\documentclass{PoS}

\usepackage[utf8x]{inputenc}
\usepackage{cite}
\usepackage{amsmath}
\usepackage{microtype}

\graphicspath{{figures/}{build/figures/}}
\DeclareMathOperator{\diag}{diag}
\newcommand{\mpl}{m_\text{Pl}}
\newcommand{\centeredgraphics}[2][]{\vcenter{\hbox{\includegraphics[#1]{#2}}}}

\title{The Gravitational Potential of Two Point Masses at Five Loops}

\ShortTitle{Gravitational Potential at Five Loops}

\author{Johannes Blümlein\\
  Deutsches Elektronen–Synchrotron, DESY,
  Platanenallee 6, D–15738 Zeuthen, Germany\\
  E-mail: \email{Johannes.Bluemlein@desy.de}}

\dedicated{
DESY 19-218\\
DO-TH 19/05\\
SAGEX-19-31
}
\author{\speaker{Andreas Maier}\\
  Deutsches Elektronen–Synchrotron, DESY,
  Platanenallee 6, D–15738 Zeuthen, Germany\\
  E-mail: \email{andreas.martin.maier@desy.de}
}

\author{Peter Marquard\\
  Deutsches Elektronen–Synchrotron, DESY,
  Platanenallee 6, D–15738 Zeuthen, Germany\\
  E-mail: \email{Peter.Marquard@desy.de}}
\abstract{
  Corrections to the Newtonian gravitational potential from general
  relativity can be derived in a combined expansion around flat
  spacetime and a small velocity of the interacting bodies. We present
  the calculation of the static five-loop corrections in an effective
  field theory framework using techniques from multi-loop computations
  in particle physics.
}

\FullConference{14th International Symposium on Radiative Corrections (RADCOR2019)\\
9-13 September 2019\\
		Palais des Papes, Avignon, France}

\begin{document}

\section{Introduction}

Accurate wave form templates are crucial for any gravitational wave
detection experiment. Significant improvements will be needed in order
to match the requirements of next-generation experiments such as LISA
or the Einstein telescope. The signals detected so far match the
expectations for gravitational wave emission from mergers of two
compact objects, such as black holes or neutron stars. Templates for
this scenario are constructed with the help of the effective one-body
formalism~\cite{Buonanno:1998gg}, which in turn relies on calculations
from first principles describing the different phases of the binary
system's evolution. Of particular importance is the inspiral phase,
where the compact objects are well separated and the relative velocity
is small compared to the speed of light.

Indeed, for a sufficiently large distance the system is described
well by Newtonian mechanics. The average kinetic and gravitational
energy are then related via the virial theorem, which predicts the scaling
\begin{equation}
  \label{eq:PN_scaling}
  v \sim \sqrt{\frac{G m}{r}} \ll 1\,,
\end{equation}
where $v$ is the characteristic velocity of a compact object, $m$ its
mass and $r$ the distance between the two objects. $G$ denotes Newton's
constant. Corrections from general relativity can then be obtained in
a simultaneous expansion in the small velocity and coupling, the
\emph{Post-Newtonian} (PN) expansion. The $k$PN order corresponds
to a suppression by $v^{2k}$.

All relevant parameters of the system are connected by the same small
expansion parameter. The system emits gravitational quadrupole waves,
whose frequency is twice the orbital frequency $\omega \sim
\tfrac{v}{r}$. The wavelength $\lambda$ is therefore of the order of
$\tfrac{r}{v}$. On the other side of the spectrum, the size of the
compact objects is characterised by the Schwarzschild radius $r_s = 2
G m \sim r v^2$. In the following, we will assume point-like objects.
Finite-size effects first contribute at 5PN order for neutron stars,
and at 6 PN order for black holes~\cite{1983LNP...124...59D}. In the events
detected at LIGO and VIRGO~\cite{LIGOScientific:2018mvr} no evidence
for non-zero spin has been observed so far and we will not discuss
spin effects in the remainder of this note.

For conservative dynamics, the Lagrangian and Hamiltonian of an
inspiraling binary system has been calculated to 4PN order in
different
formalisms~\cite{Damour:2016abl,Bernard:2016wrg,Foffa:2016rgu,Foffa:2019rdf,Foffa:2019yfl}. First
corrections at 5PN order have been obtained in~\cite{Bern:2019nnu,Foffa:2019hrb,Blumlein:2019zku,Bini:2019nra}. In the following we
discuss the calculation of the static 5PN interaction potential in a
non-relativistic effective field theory~\cite{Goldberger:2004jt}.

\section{Expansion of the general relativity action}
\label{sec:exp_SGR}

Our starting point is the Einstein-Hilbert action in harmonic gauge in
$d=3-2\epsilon$ spatial dimensions and one time dimension, viz.
\begin{equation}
  \label{eq:S_EH+GF}
  S_\text{EH} + S_\text{GF} = \frac{1}{16\pi G}\int d^{d+1}x\ \sqrt{-g}\left(R-\frac{1}{2} \Gamma_\mu \Gamma^\mu\right)\,,
\end{equation}
where $g$ is the determinant of the metric $g^{\mu\nu}$, $R$ the
scalar curvature, and $\Gamma^\mu =
g^{\alpha\beta}\Gamma^\mu_{\alpha\beta}$ with Christoffel symbols
$\Gamma^\mu_{\alpha\beta}$. Our aim is to first perform a PN expansion
of this action, which includes an expansion around the flat metric
$\eta = \diag(-1,1,1,1)$. In this sense, the compact objects induce
non-perturbative short-distance fluctuations which can be absorbed into
the point-particle action
\begin{equation}
  \label{eq:S_pp}
  S_{\text{pp}} = - \sum_{i=1}^2 m_i \int d\tau = - \sum_{i=1}^2 m_i \int dt\ \sqrt{-g_{\mu\nu} \frac{\partial x_i^\mu}{\partial t} \frac{\partial x_i^\nu}{\partial t}}\,.
\end{equation}
This is akin to the operator product expansion in QCD, where
non-perturbative field modes are absorbed into local condensates. The resulting action is
\begin{equation}
  \label{eq:S_GR}
  S_\text{GR} = S_\text{EH} + S_\text{GF} + S_{\text{pp}}\,.
\end{equation}
To facilitate the PN expansion of this action we employ a temporal
Kaluza-Klein decomposition of the metric~\cite{Kol:2010si}:
\begin{equation}
  \label{eq:g_KK}
  g^{\mu\nu} = e^{2{\phi}}
  \begin{pmatrix}
    -1 & {A_j}\\
    {A_i} &
    e^{-2 \frac{d-1}{d-2}{\phi}}(\delta_{ij} + {\sigma_{ij}})-{A_i A_j}
  \end{pmatrix}\,.
\end{equation}
The coupling of the point objects to the spatial vector field $A$ and
the tensor field $\sigma$ are suppressed by $v^1$ and $v^2$,
respectively. This greatly simplifies the Feynman rules contributing
at a given order in the PN expansion.

\section{Effective theory matching}
\label{sec:EFT_matching}

In the next step, we match the expanded general relativity action of
eq.~(\ref{eq:S_GR}) to the action of a non-relativistic
effective theory of gravity (NRGR). The derivation is similar to the one of
non-relativistic QCD (NRQCD)~\cite{Caswell:1985ui}. We distinguish between two relevant
modes of the graviton fields $\phi$, $A$, and
$\sigma$. \emph{Potential} (or orbital) gravitons have wavelengths of
the order of the orbital separation $r$. In the matching to NRGR, they
are integrated out and their effects are absorbed into interaction
potentials. The \emph{radiation} modes (called ultrasoft in NRQCD) are
associated with the emitted gravitational waves with wavelengths of
the order of $\tfrac{r}{v}$. They remain part of the effective
theory. In contrast to potential gravitons, radiation gravitons are
parametrically on-shell. Both modes have frequencies of the order of
$\tfrac{r}{v}$.

At leading order, NRGR is just Newtonian mechanics with the action
\begin{equation}
  \label{eq:S_NRGR}
  S_\text{NRGR} = \int dt\ \frac{1}{2} m_1 v_1^2 + \frac{1}{2} m_2 v_2^2 + \frac{Gm_1m_2}{r} + \dots\,,
\end{equation}
where the ellipsis denotes higher-order PN corrections to the kinetic
energy, the potential, and terms including the radiation fields.

For the matching of the two theories, i.e. to determine the parameters
of the effective theory, we equate the amplitudes for the scattering
of the two compact objects under the exchange of a four-momentum $q$
that is negligible compared to the object masses $m_1, m_2$. On the
effective theory side, the objects interact via the potential. In
perturbative general relativity, the interaction is transmitted by the
gravitons $\phi,A,\sigma$ in the potential region. The matching equation then reads
\begin{align}
  \label{eq:matching}
      &\centeredgraphics{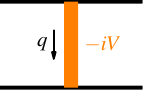} + \frac{1}{2!}\ \centeredgraphics{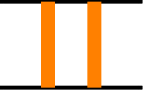} +
      \frac{1}{3!}\ \centeredgraphics{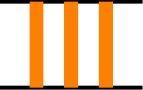} + \dots \notag\\[.5em]
      ={}&
      \centeredgraphics{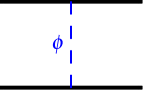} + \centeredgraphics{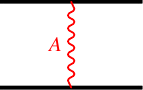}+
      \centeredgraphics{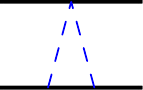} + \centeredgraphics{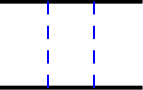} +
      \centeredgraphics{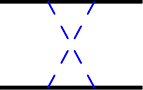} + \cdots\,,
\end{align}
where the solid black lines illustrate the classical point
object sources. Diagrams with closed graviton loops correspond to quantum
corrections. These are highly suppressed and can be safely neglected.

Adding unity and taking the logarithm on both sides of the matching
equation~\eqref{eq:matching} eliminates all higher iterations of the
potential insertion in the effective theory and the graviton exchange
diagrams that are reducible when cutting all source lines, for
example the box and crossed-box diagrams shown
in~\eqref{eq:matching}. The reason for this is the
following~\cite{Fischler:1977yf}. In position space, we label each
interaction vertex $V$ with a time $t_V$ and integrate over all time
variables. Initially, the vertices along a given source line in each
diagram are ordered in time. That is, a source line connecting two
vertices $1$ and $2$ numbered from left to right corresponds to a step
function $\Theta(t_2-t_1)$. These step functions can be eliminated by
adding up all diagrams obtained by permuting all vertices along each
source line. After removing all step functions any source-reducible
diagrams factorises and cancels against the product of lower-order
diagrams when taking the logarithm.

As an example, let us consider the one-loop seagull
diagram. Indicating the time ordering with an arrow along the source
line we can write
\begin{equation}
  \label{eq:sym_seagull}
  \underset{\Theta(t_2-t_1)}{\centeredgraphics{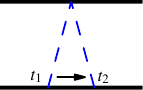}} =
  \frac{1}{2}\left(\,\underset{\Theta(t_2-t_1)}{\centeredgraphics{pot_1l_seagull_xy}} +
    \underset{\Theta(t_1-t_2)}{\centeredgraphics{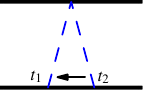}}\,\right)
  = \frac{1}{2}\ \centeredgraphics{pot_1l_seagull}\,.
\end{equation}
In the second step we have exploited the symmetry of the diagram under
exchange of the vertices on the lower source line. Applying the same
procedure to the sum of a one-loop box and the corresponding
crossed-box diagram we obtain
\begin{equation}
  \begin{split}
    &\centeredgraphics{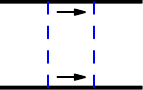} + \centeredgraphics{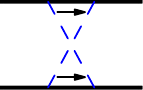} =
    \centeredgraphics{pot_1l_box_xy} + \centeredgraphics{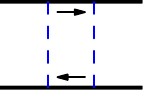}\\[0.5em]
      ={}&\frac{1}{2}\left(
        \ \centeredgraphics{pot_1l_box_xy} + \centeredgraphics{pot_1l_box_yx}
        + \centeredgraphics{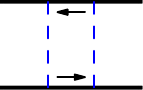} + \centeredgraphics{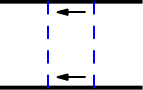}
        \ \right)\\[0.5em]
      ={}&\frac{1}{2}\ \centeredgraphics{pot_1l_box} = \frac{1}{2} \left(\
        \centeredgraphics{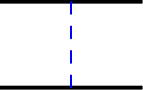}\ \right)^2\,.
  \end{split}
\end{equation}
The factorisation in the last step is evident in position space. The
resulting product is cancelled against the contribution from the
square of the tree-level scalar exchange diagram upon taking the
logarithm.

In the following, we restrict ourselves to the static limit
$v_1=v_2=0$. In this case, the sources only couple directly to the
scalar $\phi$. Vector gravitons $A$ are then always produced in pairs
and are therefore always part of pure graviton loops. As mentioned
before, pure graviton loops do not contribute in the classical limit
and the $A$ field decouples completely from the theory.

\section{Setup of the calculation}
\label{sec:setup}

The diagrams contributing to the static gravitational potential can be
calculated using standard multi-loop tools and techniques. We employ
\texttt{QGRAF}~\cite{Nogueira:1991ex} to generate
\texttt{FORM}~\cite{Vermaseren:2000nd, Tentyukov:2007mu} code for the
diagrams. We eliminate diagrams that are irrelevant according to the
criteria listed in Section~\ref{sec:EFT_matching}. Since the only
scale is given by the external momentum $q$ the diagrams belong to the
well-studied class of massless propagators. At five-loop order, we
identify 22 topologies, see~\cite{Blumlein:2019zku}.

In the next step, we perform the symmetrisation discussed in
Section~\ref{sec:EFT_matching} and insert the momentum-space Feynman
rules obtained from the expanded general relativity action,
cf. Section~\ref{sec:exp_SGR}:
\begin{align}
      \centeredgraphics[width=40pt]{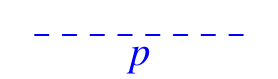}={}& - \frac{i}{2c_d\vec{p\,}^2}\,,\displaybreak[0]\\
  \centeredgraphics[width=40pt]{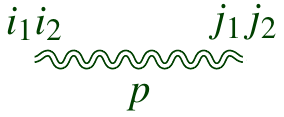}={}& -
  \frac{i}{2\vec{p\,}^2}\big(\delta_{i_1j_1}\delta_{i_2j_2}+\delta_{i_1j_2}\delta_{i_2j_1}
                          + (2-c_d)\delta_{i_1i_2}\delta_{j_1j_2}\big)\,,\displaybreak[0]\\
\centeredgraphics[width=40pt]{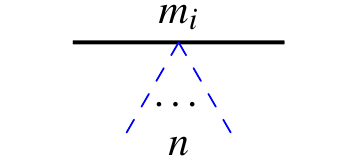} ={}& -i\frac{m_i}{\mpl^n}\,,\displaybreak[0]\\
  \centeredgraphics[width=30pt]{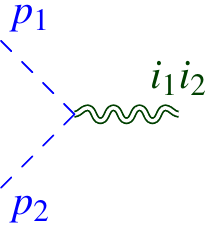} ={}&i\frac{c_d}{2\mpl}(V_{\phi\phi\sigma}^{i_1 i_2} +
V_{\phi\phi\sigma}^{i_2 i_1}) \,,\displaybreak[0]\\
  V_{\phi\phi\sigma}^{i_1 i_2} ={}&
\vec{p}_1\cdot \vec{p}_2 \delta^{i_1i_2}- 2 p_1^{i_1} p_2^{i_2}\,,\displaybreak[0]\\[1em]
 \centeredgraphics[width=30pt]{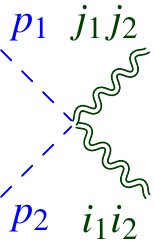} ={}&
i\frac{c_d}{16\mpl^2}(V_{\phi\phi\sigma\sigma}^{i_1i_2,j_1j_2} +
V_{\phi\phi\sigma\sigma}^{i_2i_1,j_1j_2} +
V_{\phi\phi\sigma\sigma}^{i_1i_2,j_2j_1} +
V_{\phi\phi\sigma\sigma}^{i_2i_1,j_2j_1})\,,\displaybreak[0]\\
V_{\phi\phi\sigma\sigma}^{i_1i_2,j_1j_2} ={}& \vec{p}_1\cdot \vec{p}_2
(\delta^{i_1i_2}\delta^{j_1j_2} - 2\delta^{i_1j_1}\delta^{i_2j_2})
%\notag\\
%&
-2(p_1^{i_1}p_2^{i_2}\delta^{j_1j_2} + p_1^{j_1}p_2^{j_2}\delta^{i_1i_2})
   + 8\delta^{i_1j_1} p_1^{i_2}p_2^{j_2}\,,\displaybreak[0]\\
\centeredgraphics[width=40pt]{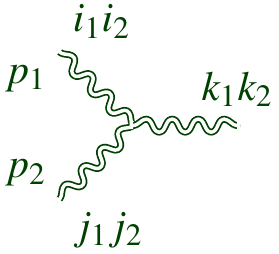} ={}&\frac{i}{32\mpl}(\tilde{V}_{\sigma\sigma\sigma}^{i_1i_2,j_1j_2,k_1k_2} +
       \tilde{V}_{\sigma\sigma\sigma}^{i_2i_1,j_1j_2,k_1k_2})\,,\displaybreak[0]\\
  \tilde{V}_{\sigma\sigma\sigma}^{i_1i_2,j_1j_2,k_1k_2} ={}&
       V_{\sigma\sigma\sigma}^{i_1i_2,j_1j_2,k_1k_2} +
       V_{\sigma\sigma\sigma}^{i_1i_2,j_2j_1,k_1k_2} +
       V_{\sigma\sigma\sigma}^{i_1i_2,j_1j_2,k_2k_1} +
  V_{\sigma\sigma\sigma}^{i_1i_2,j_2j_1,k_2k_1}
\,,\displaybreak[0]\\
V_{\sigma\sigma\sigma}^{i_1i_2,j_1j_2,k_1k_2}
={}&(\vec{p}_1^2+\vec{p}_1\cdot\vec{p}_2+\vec{p}_2^2)\*\Big(-\delta^{j_1j_2}\*\big(2\*\delta^{i_1k_1}\*
\delta^{i_2k_2}-\delta^{i_1i_2}\*\delta^{k_1k_2}\big)
\notag\\
&\quad+2\*\big[\delta^{i_1j_1}\*\big(4\*\delta^{i_2k_1}\*\delta^{j_2k_2}-\delta^{i_2j_2}\*\delta^{k_1k_2}\big)
-\delta^{i_1i_2}\*\delta^{j_1k_1}\*\delta^{j_2k_2}\big] \Big)
\notag\\
&+2\*\Big\{4\*\big(p_1^{k_2}\*p_2^{i_2}-p_1^{i_2}\*p_2^{k_2}\big)\*\delta^{i_1j_1}\*\delta^{j_2k_1}
\notag\\
&\quad+2\*\big[\big(p_1^{i_1}+p_2^{i_1}\big)\*p_2^{i_2}\*\delta^{j_1k_1}\*\delta^{j_2k_2}-p_1^{k_1}\*p_2^{k_2}\*
\delta^{i_1j_1}\*\delta^{i_2j_2}\big]
\notag\\
&\quad+\delta^{j_1j_2}\*\big[p_1^{k_1}\*p_2^{k_2}\*\delta^{i_1i_2}+2\*\big(p_1^{k_2}\*p_2^{i_2}-p_1^{i_2}\*
p_2^{k_2}\big)\*\delta^{i_1k_1}-\big(p_1^{i_1}+p_2^{i_1}\big)\*p_2^{i_2}\*\delta^{k_1k_2}\big]
\notag\\
&\quad+p_2^{j_2}\*\Big(4\*p_1^{i_2}\*\delta^{i_1k_1}\*\delta^{j_1k_2}+p_1^{j_1}\*\big(2\*\delta^{i_1k_1}\*
\delta^{i_2k_2}-\delta^{i_1i_2}\*\delta^{k_1k_2}\big)
\notag\\
&\qquad+2\*\big[\delta^{i_1j_1}\*\big(p_1^{i_2}\*\delta^{k_1k_2}-2\*p_1^{k_2}\*\delta^{i_2k_1}\big)-p_1^{k_2}\*
\delta^{i_1i_2}\*\delta^{j_1k_1}\big] \Big)
\notag\\
&\quad+p_1^{j_2}\*\Big(p_1^{j_1}\*\big(2\*\delta^{i_1k_1}\*\delta^{i_2k_2}-\delta^{i_1i_2}\*\delta^{k_1k_2}\big)
-4\*p_2^{i_2}\*\delta^{i_1k_1}\*\delta^{j_1k_2}
\notag\\
&\qquad+2\*\big[p_2^{k_2}\*\delta^{i_1i_2}\*\delta^{j_1k_1}+\delta^{i_1j_1}\*\big(2\*p_2^{k_2}\*\delta^{i_2k_1}
-p_2^{i_2}\*\delta^{k_1k_2}\big)\big]
\Big)\Big\}\,,
\end{align}
with $c_d = 2\tfrac{d-1}{d-2}, \mpl = 1/\sqrt{32 \pi G}$. There is no
propagator associated with the classical sources. We perform an
expansion around $\epsilon = \tfrac{3-d}{2} \to 0$ and exploit
integration-by-parts identities~\cite{Chetyrkin:1981qh} to reduce the
scalar integrals to a small set of master integrals. The reduction is
based on Laporta's algorithm~\cite{Laporta:2001dd}, implemented in our
in-house code \texttt{crusher}~\cite{crusher}. We find that at five loops only four
out of eight master integrals contribute in the limit $\epsilon \to
0$. Similar to the findings of~\cite{Foffa:2019hrb}, we observe that
all of them factorise into lower-order integrals. Defining $L-$loop
master integrals with propagator momenta $\vec{p}_1,\dots,\vec{p}_P$
as
\begin{equation}
  \label{eq:M_L-loop}
  M_{P} = \int \left(\prod_{i=1}^L \frac{d^dl_i}{\pi^{d/2}}\right)\frac{1}{\vec{p}_1^2\dots \vec{p}_P^2}\,,
\end{equation}
we obtain
\begin{align}
\centeredgraphics{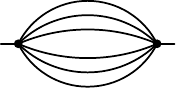} ={}&  \frac{\Gamma\left(6 - \frac{5 d}{2}\right) \Gamma^6\left(-1 +
\frac{d}{2}\right)}{\Gamma(-6 + 3 d)}\,,\\
\centeredgraphics{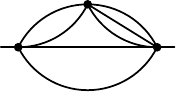} ={}&
\frac{\Gamma \left(
        7-\frac{5 d}{2}\right) \Gamma \left(3-d\right) \Gamma\left(
        2-\frac{d}{2}\right) \Gamma^7 \left(
        -1+\frac{d}{2}\right) \Gamma(5-2d)}{\Gamma \left(5-\frac{3}{2} d\right) \Gamma(-2+d) \Gamma \left(
        -3 + \frac{3}{2} d\right) \Gamma \left(-7+3 d\right)}\,,\\
\centeredgraphics{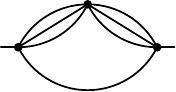} ={}&
\frac{\Gamma \left(
        7-\frac{5 d}{2}\right) \Gamma^2 (3-d) \Gamma^7 \left(
        -1+\frac{d}{2}\right) \Gamma \left(
        -6+\frac{5 d}{2}\right)}{\Gamma \left(6-2 d\right) \Gamma^2 \left(
        -3+\frac{3 d}{2}\right) \Gamma \left(-7+3 d\right)}\,,
\\
\centeredgraphics{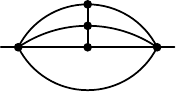} ={}& 6 \pi^{7/2} S_\epsilon^5 \Biggl[\frac{2}{\epsilon} - 4(1 - \ln(2)) - \left(48 + 8 \ln(2) - 4
\ln^2(2) - 105 \zeta_2\right)
\epsilon
+ \Biggl(480 - 96 \ln(2)
\nonumber\\ &
- 8 \ln^2(2)
+ \frac{8}{3} \ln^3(2) - 530 \zeta_2 + 402 \ln(2) \zeta_2 - \frac{1522}{3}
\zeta_3\Biggr) \epsilon^2 \Biggr] + O(\epsilon^3)\,,
\end{align}
where $S_\epsilon = \exp(-\gamma_E \epsilon)$. The first three
diagrams can be decomposed into products of one-loop integrals
\begin{equation}
\centeredgraphics{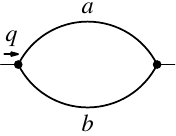} = \frac{1}{(q^2)^{a+b-d/2}} \frac{
\Gamma\left(\frac{d}{2}-a\right)
\Gamma\left(\frac{d}{2}-b\right)
\Gamma\left(a+b-\frac{d}{2}\right)}{\Gamma(a) \Gamma(b)
\Gamma\left(d-a-b\right)}\,,
\end{equation}
whereas the last master integral also contains a factor of
\begin{equation}
  \centeredgraphics{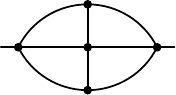}=
  2 \pi^2 S_\epsilon^4 \Biggl[\frac{1}{\epsilon^2} + \frac{2}{\epsilon} - 2(16 -  \zeta_2) + 16 \Biggl[9 - 6 \zeta_2
  \left(\frac{13}{8} -
    \ln(2) \right)
  - \frac{77}{6} \zeta_3\Biggr] \Biggr]\epsilon + O(\epsilon^2)\,,
\end{equation}
which was calculated in~\cite{Lee:2015eva,Damour:2017ced}.

Up to four loops, our results for the potential agree with the
previous calculations in the effective field theory
framework~\cite{Foffa:2016rgu,Goldberger:2004jt,Gilmore:2008gq,Foffa:2011ub}. At
five loops we obtain
\begin{equation}
  \label{eq:result}
  V_{\text{5PN}}^S =
  \frac{G^6}{r^6} m_1 m_2 \Biggl[
  \frac{5}{16} (m_1^5+m_2^5)
  + \frac{91}{6} m_1 m_2 (m_1^3+m_2^3)
  + \frac{653}{6} m_1^2 m_2^2 (m_1 + m_2)
  \Biggr]\,,
\end{equation}
in agreement with a concurrent independent calculation~\cite{Foffa:2019hrb}.

\subsection*{Acknowledgements}

This project has received funding from the European Union's Horizon
2020 research and innovation programme under the Marie
Skłodowska-Curie grant agreement No. 764850, SAGEX, and COST
action CA16201: Unraveling new physics at the LHC through the
precision frontier.

\bibliographystyle{JHEP}
\bibliography{biblio.bib}

% \begin{thebibliography}{99}

% \end{thebibliography}

\end{document}